\documentclass[letter]{aa} % for the letters
\usepackage{graphicx}
\usepackage{txfonts}
\begin{document}

   \title{The mixing of dust and gas in the high latitude translucent cloud MBM 40}

   \subtitle{}

   \author{Marco Monaci
          \inst{1}
          \and
          Loris Magnani
          \inst{2}
          \and
          Steven N.Shore
          \inst{1}
          }

   \institute{Dipartimento di Fisica, Universit\`a di Pisa, Largo Bruno Pontecorvo 3, Pisa \\
              \email{monaci93@gmail.com} \\
            \email{steven.neil.shore@unipi.it}
         \and
             Department of Physics and Astronomy, University of Georgia, Athens, GA 30602-2451 \\
             \email{loris@uga.edu}
             }

   \date{Received 20 September 2022 ; accepted 25 October 2022}

% \abstract{}{}{}{}{}
% 5 {} token are mandatory
 
  \abstract
  % context heading (optional)
  % {} leave it empty if necessary  
   {High latitude molecular clouds (hereafter HLMCs) permit the study of interstellar gas dynamics and astrochemistry with good accuracy due to their proximity, generally clear lines of sight, and lack of internal star-forming activity which can heavily modify the physical context. MBM 40, one of the nearest  HLMCs, has been extensively studied, making it a superb target to infer and study the dust-to-gas mixing ratio (DGMR).}
  % aims heading (mandatory)
   {The mixing of dust and gas in the interstellar medium remains a fundamental issue to keep track of astrochemistry evolution and molecular abundances. Accounting for both molecular and atomic gas is difficult because $\mathrm{H_2}$ is not directly observable and \ion{H}{i} spectra always show different dynamical profiles blended together which are not directly correlated with the cloud. We used two  independent strategies to infer the molecular and atomic gas column densities and compute the dust-to-gas mixing ratio.}
  % methods heading (mandatory)
{We combined  \ion{H}{i} 21 cm and \element[][12]CO line observations with the IRAS 100 $\mathrm{\mu m}$ image to infer the dust-to-gas mixing ratio within the cloud. The cloud 21 cm profile was extracted using a hybrid Gaussian decomposition where \element[][12]CO was used to deduce the total molecular hydrogen column density. Infrared images were used to calculate the dust emission.} 
  % results heading (mandatory)
   {The dust-to-gas mixing ratio is nearly uniform within the cloud as outlined by the hairpin structure. 
   %The bulk of the cloud seen in \element[][12]CO and \element[][13]CO is embedded where the mixing ratio is below $1.5 \cdot 10^{20} \ \mathrm{MJy \ sr^{-1} \ cm^{2}}$. Using the cloud distance of 93 pc the east and west ridges of the cloud structure are 0.89 pc and 1 pc wide respectively, meaning that the mixing is nearly constant on such physical scale. Lastly,.
   The total hydrogen column density and 100 $\mathrm{\mu m}$ emissivity are linearly correlated over a range in $\mathrm{N(H_{tot})}$ of one order of magnitude.}
   {}
  
    %{\bf The method used in this work allows to estimate the dust to gas ratio with three nearly independent observations, the 21 cm hydrogen line for atomic hydrogen, \element[][12]CO for molecular hydrogen and 100 $\mathrm{\mu m}$ infrared images for dust component. 
    %{\bf \color{red} THOUGHTS - SORRY FOR TOO MANY MIGHT} \\
    %{ \bf  The constancy of the gas to dust ratio across the molecular structure indicates that gas and dust are well mixed regardless the total gas density: if the mixing mechanism is turbulence, then the mean grain size must be small; consequently turbulence might act like a sorting mechanism for grain size distribution; H$_2$ high efficiency formation on grain surfaces might indicate that H$_2$ is created mainly on small grains and turbulence might help molecular formation providing a well mixed grain ensamble. Furthermore, molecules formation supplies energy that might be reirradiated by dust, keeping cool the molecular structure (alongside CO - but where is the spectral signature, if any?).} 

        % maybe we need other keywords here

   \maketitle
%
%________________________________________________________________
\section{Introduction}\label{sec:intro}
The admixture of dust and gas is an essential, but poorly determined, property of the interstellar medium.  It affects the modeling of radiative balance, moderates  astrochemical processes, and affects star formation within molecular clouds (for the theoretical aspect see, e.g., \cite{lee2017dynamics}, \cite{tricco2017dust}, \cite{marchand2021fast}; for the observational picture, e.g., \cite{reach2017variations}). The widely adopted value for the mass ratio in our Galaxy is $\sim100-150$ (\cite{hildebrand1983determination}), but it varies in other  galaxies (\cite{young1986co}) and even in clouds in our Galaxy (\cite{reach2015variations}).\footnote{We must emphasize at the beginning of this Letter that our aim is to study the mixing of the components and not the mass ratio, as conventionally discussed, because we are trying to avoid the uncertainties induced by adopting specific dust models.}

The usual procedure to obtain the dust-to-gas ratio (hereafter DGR) infers the neutral hydrogen column densities from the dust extinction which is then used to obtain the dust mass.  Alternatively, atomic resonance transitions have been used to infer the atomic column densities, but these are altered by density-dependent depletion in the gas phase and from the spotty sampling of any intervening clouds because of the sparse coverage by stars and extragalactic sources.  The sparseness of background sources also  complicates the distance determination of clouds using standard methods such as star counting, Wolf diagrams, or multiband extinction measurements (e.g., \cite{sun2021extinction}, \cite{lv2018gas}, \cite{liljestrom1988neutral}). These techniques are hampered by the proximity of the translucent and high latitude molecular clouds (see, e.g., \cite{magnani2017dirty}).  Moreover, different lines of sight through these clouds may sample very different structures (e.g., \cite{lombardi2014herschel}) .  If the DGRs differ, this complicates extinction corrections and inferences about the cloud masses based on infrared imaging.  It is now possible, however, to obtain the atomic column densities directly following the completion of all-sky \ion{H}{i} surveys that complement the infrared maps obtained for Galactic and cosmological surveys.  

The object of our study, MBM 40, is a small molecular structure embedded in a larger \ion{H}{i} flow (\cite{shore2003}). The molecular gas distribution has been discussed extensively (\cite{magnani1985molecular}, MBM; \cite{lee2002co}; \cite{shore2003}, SMLM; \cite{shore2006}, SLCM; \cite{chastain2009high}, CCM10).  The CO(1-0) molecular gas distribution is complex with the denser gas in a hairpin structure surrounded by a diffuse envelope.  The distance to the cloud is 93 pc (\cite{zucker2019}) and previous studies (\cite{magnani1996catalog}; SMLM; CCM10) have derived a molecular mass of 20 - 40 \ M$_\sun$.  There is no evidence of star formation despite a rigorous search (\cite{magnani1996search}).  This cloud provides an  exemplar of a non-star-forming medium where no internal processing has  affected the dust properties.  We should, therefore, see a more nearly pristine presentation of the DGR and its uniformity than would be obtained from a more active source.

%-----------------------------------------------------------------------------
\section{Data}\label{sec:data}
We have used a range of archival data for this study.  We briefly describe here the individual data sets.

\subsection{GALFA}\label{subsec:galfa}
\begin{figure}[h!]
\centering
        \includegraphics[width=\hsize]{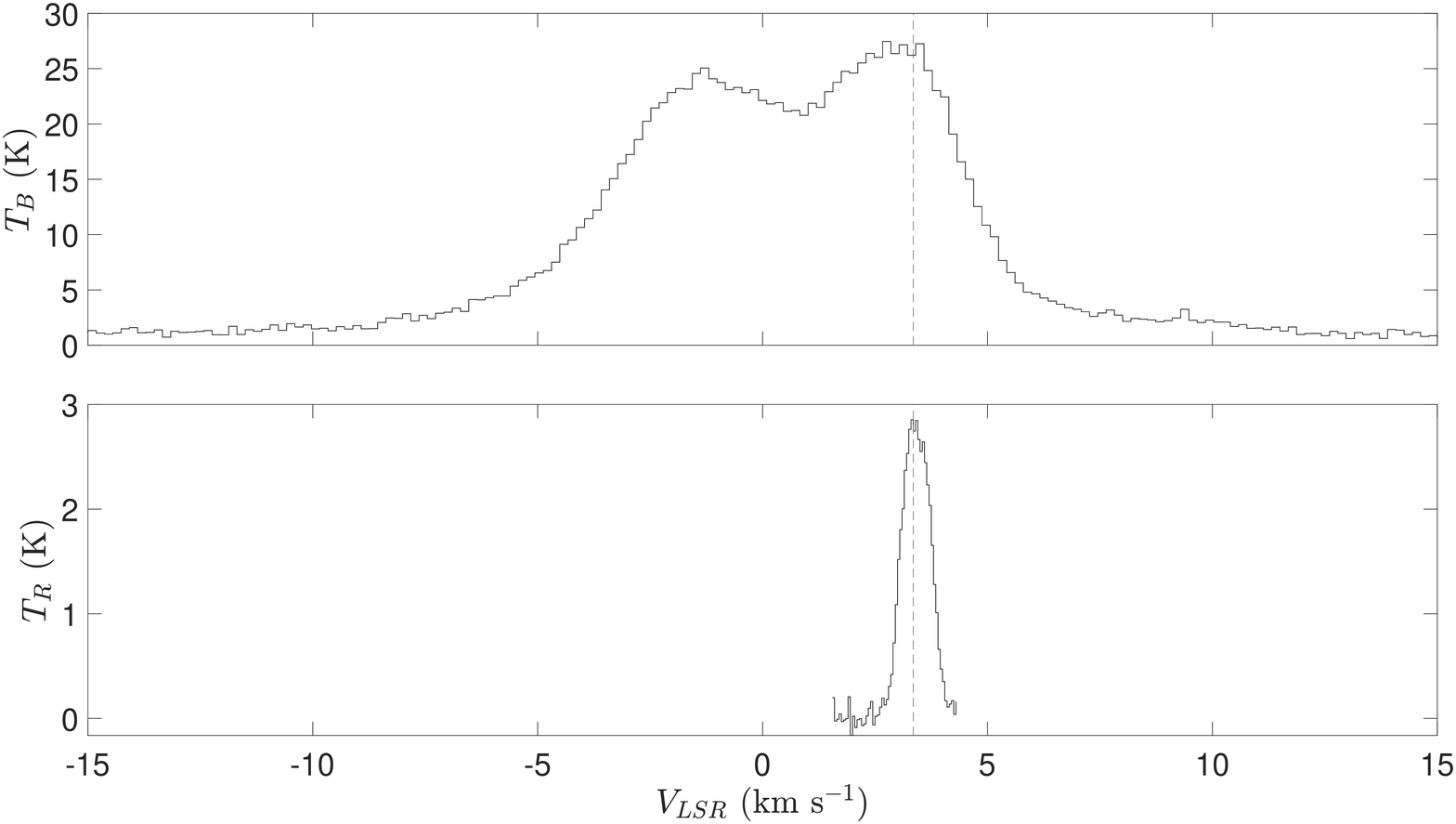}
    \caption{ Sample spectra of MBM 40 used in this study.  \textit{(Top panel.)} Single GALFA spectrum  at $\alpha=242.7583\degr , \delta=+21.8084\degr$ (J2000.0), in the lower part of the western ridge of MBM 40. \textit{(Bottom panel.)} Sample FCRAO \element[][12]CO spectrum near the same position as the \ion{H}{i}, obtained by averaging 64 spectra in an $8 \times 8$ pattern to enhance the S/N. The vertical dashed line shared by two plots indicates the bulk velocity of the cloud ($3.35 \ \mathrm{ km \ s^{-1}}$). The \ion{H}{i}  spectrum shows the brightness temperature ($T_\mathrm{B}$), and the \element[][12]CO spectrum is in terms of the radiation temperature ($T_\mathrm{R}$).}
        \label{fig:galfa_fcrao}
\end{figure}

The Galactic Arecibo L-Band Feed Array HI (GALFA-HI, \cite{peek2011galfa},) is an extended  survey between $-1 \degr \lesssim \delta \lesssim 38 \degr$ with an angular resolution of $4\arcmin$ and a $0.184 \ \mathrm{km \ s^{-1}}$ spectral resolution using the William E. Gordon 305-m telescope at the Arecibo Observatory (\cite{peek2018}).\footnote{During these observations the Arecibo Observatory was operated by SRI International under a cooperative agreement with the National Science Foundation, and in alliance with Ana G. M\'endez-Universidad Metropolitana, and the Universities Space Research Association.}.  We used the narrow bandwidth data repository (see \cite{galfadata2017}) centered at $\mathrm{0 \; km \ s^{-1}}$; each spectrum has $2 \, 048$ channels spanning  $\left| v_{\mathrm{LSR}} \right| \lesssim 188 \; \mathrm{km \ s^{-1}} $. We used ancillary data furnished with the GALFA datacube to correct for stray radiation and cropped the PPV datacube in  position to match the extension and velocity ($\left| v_{\mathrm{LSR}} \right| \lesssim 15 \  \mathrm{km \ s^{-1}}$) of  MBM 40, which is detected between $\sim 2$ and $\sim 4 \ \mathrm{km \ s^{-1}}$ in molecular gas.  
The top panel of Fig. \ref{fig:galfa_fcrao} shows an example of a GALFA narrowband spectrum (within the range $\vert{v_{\mathrm{LSR}}}\vert \leq 20 \ \mathrm{km \ s^{-1}}$) with a signal-to-noise ratio (S/N) $\sim 60$; each spectrum within the cloud has a similar S/N. 

\subsection{FCRAO}\label{subsec:fcrao}
The \element[][12]CO observations were obtained using the Five College Radio Astronomy Observatory (FCRAO\footnote {The FCRAO was supported in part by the National Science Foundation and was operated with the permission of the Metropolitan District Commission, Commonwealth of Massachusetts.}) in early 2000. The full datacube is composed of $24 \, 576$ frequency-switched spectra with a velocity resolution of about $0.05 \ \mathrm{km \ s^{-1}}$.  We  used only the 56 central channels centered at $3 \ \mathrm{km \ s^{-1}}$. The average rms noise value is $0.7 \ \mathrm{K}$ (\cite{shore2003}). The \element[][12]CO radiation temperature ($T_\mathrm{R}$) was calculated taking the antenna temperature and then dividing by $\eta_{\mathrm{fss}}\eta_{\mathrm{c}}$, where $\eta_{\mathrm{fss}}$ is the forward-scattering and spillover efficiency ($\simeq 0.7$) and $\eta_{\mathrm{c}}$ is the source filling factor, which we assumed to be unity (see SMLM for a complete discussion of FCRAO observations).
The bottom panel of Fig. \ref{fig:galfa_fcrao} shows the equivalent FCRAO \element[][12]CO profile near the same position of the \ion{H}{i} GALFA spectrum. The \ion{H}{i} profile shows different components, one of which is compatible with the MBM 40 bulk velocity of $\sim 3 \ \mathrm{km \ s^{-1}}$.

\begin{figure}[h!]
\centering
        \includegraphics[width=\hsize]{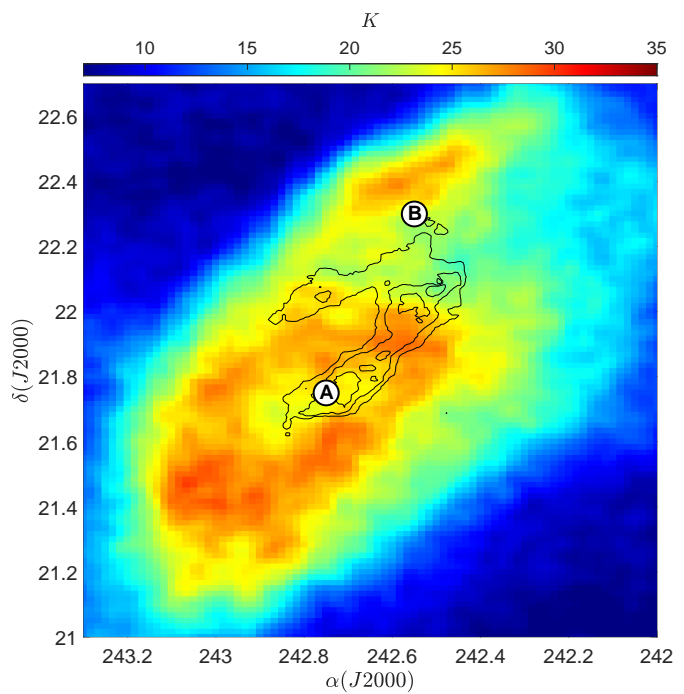}
    \caption{\ion{H}{i} brightness temperature at $3.4 \; \mathrm{km \ s^{-1}}$ with integrated line \element[][12]CO intensity (black contours at levels 2, 3, and 4 $\mathrm{K \ km \ s^{-1}}$). White circles denote the positions to which we refer in the text. In this velocity slice (very near the molecular cloud bulk velocity), the \ion{H}{i} emission and \element[][12]CO emission (position A) are anticorrelated.}
        \label{fig:positions}
\end{figure}

The HI and \element[][12]CO temperature contours for a restricted velocity range are shown in Fig. \ref{fig:positions}. Here we identify two positions to which we refer in the following sections. 
%{\color{red} Question: do we actually refer to these positions by these names in the following sections - if not, then we can get rid of the following long sentence}. Position 1 shows in \element[][12]CO a very high spatial correlation;
%position 2 shows narrower profiles and a narrowing in molecular gas spatial distribution; position 3 is near the cloud boundary and  shows weak \element[][12]CO and virtually no \element[][13]CO  emission; position 4  shows double peak profiles in both \element[][12]CO and \element[][13]CO.

\subsection{IRAS 100 $\mu$m \ dust \ map}\label{subsec:iras}
For the dust distribution, we used the IRAS 100 $\mu$m IRIS images that have  a spatial resolution of about $2\arcmin$ (\cite{iris2005}).  These were published after our first study of MBM 40 and have reduced striping, improved zodiacal light subtraction, and a zero level compatible with DIRBE. The image is an interpolated $50 \times 50$ pixel matrix with the same \ion{H}{i} and \element[][12]CO spatial resolution centered at $RA = 16^\mathrm{h} 10^\mathrm{m}57\fs 19$, $DEC = +21\degr 52\arcmin 29\farcs 28$. We did not use Planck data for the principal study because the spatial resolution is slightly lower and further reduced by a required  interpolation to a common coordinate grid (from galactic to equatorial).  We show in the Appendix, however, that our conclusions are unchanged based on Planck.

%------------------------------------------------------------------------------
\section{Data analysis}\label{sec:analysis}

\subsection{Velocity slices}\label{subsec:velslice}
   \begin{figure*}
   \resizebox{\hsize}{!}
            {\includegraphics[bb=10 10 1100 650,clip]{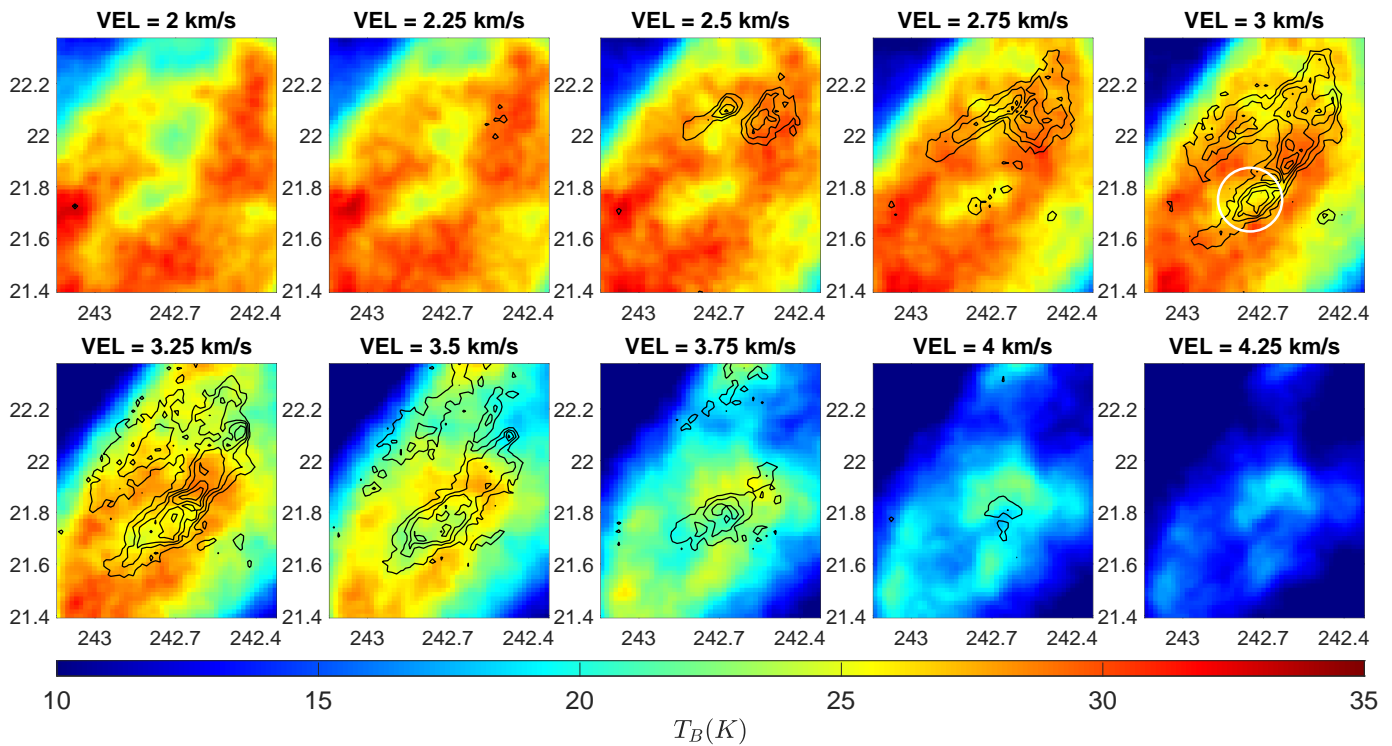}}
      \caption{\ion{H}{i} velocity slices (color scale) compared to the same \element[][12]CO velocity slices (black contours). Contours span from $1 \ \mathrm{K}$ to $5 \ \mathrm{K}$ in steps of $1 \ \mathrm{K}$. The white circle in the $3 \ \mathrm{km \ s^{-1}}$ panel marks position A in the western ridge.
              }
         \label{fig:galfa_fcrao_slice}
   \end{figure*}
   
   The comparison between \element[][12]CO and \ion{H}{i} profiles in different velocity slices is shown in Fig. \ref{fig:galfa_fcrao_slice}. Because the spectral resolution of \ion{H}{i} is coarser than \element[][12]CO, a linear interpolation was performed to match the velocity resolution of the FCRAO data (see the next subsection for further details). The \ion{H}{i} is more extended in velocity than the molecular gas traced by \element[][12]CO.  It  appears as a "cocoon" (\cite{shore2003},\cite{verschuur1974}) around cloud molecular gas and shows some internal structures. We note that \ion{H}{i} is present at both lower and higher velocities (i.e., $v < 2.25 \ \mathrm{km \ s^{-1}}$ or $ v > 4 \ \mathrm{km \ s^{-1}}$) where no \element[][12]CO is present, tracing foreground and background gas. 

The \ion{H}{i} spatial resolution is sufficient to compare the atomic and molecular gas structures mapped by \element[][12]CO. The velocity slices show an anticorrelation between atomic and molecular gas, especially at position A, marked by a white circle in the $3 \ \mathrm{km \ s^{-1}}$ slice in Fig. \ref{fig:galfa_fcrao_slice}.  The \element[][12]CO enhancement is associated with a lower atomic gas column density, indicative of a phase transition.  This condensation is visible in all velocity slices.

\subsection{Data rebinning and interpolation}\label{subsec:rebinning}
The 21 cm and \element[][12]CO maps differ in spectral and spatial resolutions. To compare these data, spectral and spatial linear interpolations were performed. Each \ion{H}{i} spectrum was linearly interpolated using the \element[][12]CO FCRAO velocities ($\Delta v = 0.05 \ \mathrm{km \ s^{-1}}$) as query points. The final spectral \ion{H}{i} resolution is $~3.5$ times higher than the original GALFA resolution. Each \ion{H}{i} spectrum has been checked for artifacts and, although linear interpolation is sensitive to S/N, in our case S/N is over 60 for each spectrum. We then  performed a 2D linear interpolation for both \ion{H}{i} and FCRAO data to a common $50 \times 50$ pixel grid, covering about $0\fdg 8$ in RA and $1\degr$ in DEC.  Consequently, using this procedure at each position yields  linked GALFA and FCRAO spectra with the same velocity resolution.

\subsection{\ion{H}{i} profile decomposition}\label{subsec:decomp}
Each \ion{H}{i} profile is a blend of multiple components. Because the 21 cm  transition is almost always optically thin for lines of sight toward high-latitude clouds, the full line of sight contributes to the observed profile, not only to  gas associated with MBM 40, and significant effort was made to decompose each HI spectrum into simpler Gaussian components (see \cite{murray2021mach}, \cite{lindner2015autonomous}, \cite{pingel2013characterizing}). In this work we effected a two-step decomposition of each profile guided by velocity information from \element[][12]CO as explained below.  We used Gaussian profiles to separate the foreground and background as well as cloud neutral hydrogen. The \element[][12]CO observations constrain a velocity window in which the cloud is embedded, so it is possible to use this information to trim the \ion{H}{i} spectra. We emphasize that the choice of Gaussian fitting was used only to select that gas that is connected to the cloud.  This is different from recent Gaussian decomposition studies aimed at characterizing the fine structure of the cold neutral medium (e.g., \cite{murray2021mach}).

\begin{figure}[h!]
\centering
        \includegraphics[width=\hsize]{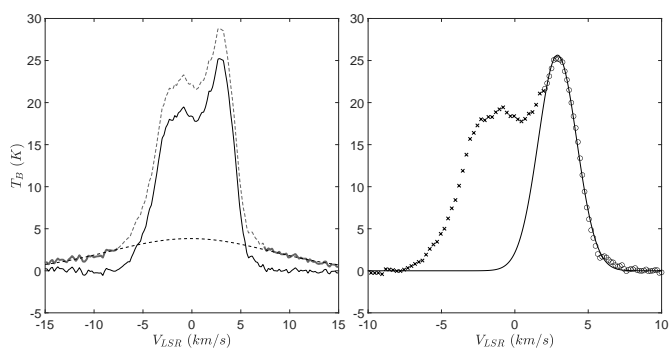}
    \caption{\textit{(Left panel.)} Subtraction of extremely diffuse gas from \ion{H}{i} sample profile. Bold gray lines are wings fitted by a single Gaussian (black dashed line) and the gray dashed line is the section we excluded from the fit; the solid black line is the result of the subtraction, where the diffuse component is severely reduced. \textit{(Right panel.)} Narrow component extraction: crosses indicate points below $2 \ \mathrm{km \ s^{-1}}$ which we suppose are not directly linked with gas traced by \element[][12]CO and which were excluded for the narrow Gaussian fit (black solid line). Circles denote points we used for the fit.}
        \label{fig:decomposition}
\end{figure}

We started the \ion{H}{i} decomposition by subtracting a very broad, diffuse component (see, \cite{verschuur1994nature}) from each GALFA profile, fitting  a Gaussian only to the wings of the profile outside the interval $-7.55\leq v_{\mathrm{LSR}} \leq 7.35 \ \mathrm{km \ s^{-1}}$, and then we fit a second Gaussian to the residual emission.  Fig. \ref{fig:decomposition} (left panel) shows an example of the procedure.

 We excluded positions where there is no clear double profile (i.e., where only diffuse gas is present in the neighborhood of MBM 40), and we fit only the redward wing of each \ion{H}{i} residual emission, avoiding  points below $2 \ \mathrm{km \ s^{-1}}$ (see Fig. \ref{fig:decomposition}, right panel). 
 This procedure cannot map all the gas because we do not know the parent distribution; however, the narrow Gaussian component is at least proportional to the total gas connected with MBM 40.

\subsection{Molecular and atomic gas column density}\label{subsec:columnden}

If \ion{H}{i} is optically thin, as in MBM 40, and if we assume that  one temperature dominates each velocity channel, then the total atomic hydrogen column density is given by the following (see for example \cite{draine2010physics}):

\begin{equation}
N(\ion{H}{i}) = 1.813 \cdot 10^{18} \int_{-\infty}^{+\infty} T_{\mathrm{B,\ion{H}{i}}}(v) \ \mathrm{d}v \ \ \ \ \ \mathrm{cm^{-2}}
\label{eq:nhi}
,\end{equation}

\noindent
where $T_{\mathrm{B,\ion{H}{i}}}(v)$ is the  brightness temperature in K and $v$ is the velocity in $\mathrm{km \ s^{-1}}$.  

The  molecular hydrogen column density, $N(\mathrm{H_2})$,  was obtained using the integrated \element[][12]CO line in $T_\mathrm{R}$  multiplied by a conversion factor $(X_\mathrm{CO}$ in units of $\mathrm{cm^{-2} \ K^{-1} \ km^{-1} \ s})$:

\begin{equation}
N(\mathrm{H_2}) = X_{\mathrm{CO}}W(\mathrm{CO_{J=1\to 0}}) \ \ \ \ \ \mathrm{cm^{-2}}
\label{eq:xco}
,\end{equation}

\noindent
where W(CO) is the velocity-integrated \element[][12]CO radiation temperature.

The value of $X_{\mathrm{CO}}$ is not constant in the Galaxy or even inside the same cloud (\cite{bolatto2013co}), and it is a source of systematic uncertainty. \cite{cotten2013} found that for MBM 40, the $X_{\mathrm{CO}}$ factor spans from $(0.6 - 3.3) \cdot 10^{20}$ with an average of $1.3 \cdot 10^{20}$: we adopted this value as well as a systematic uncertainty of about a factor of two. For each FCRAO position, we evaluated $N(\mathrm{H_2})$ as follows:

\begin{equation}
N(\mathrm{H_2}) = 1.3 \cdot 10^{20} \ \mathrm{K^{-1} \ km^{-1} \ s}  \int_{-\infty}^{+\infty} T_{\mathrm{R,CO}}(v) \ \mathrm{d}v \ \ \ \ \ \mathrm{cm^{-2}}
\label{eq:nh2}
,\end{equation}

\noindent
where $T_{\mathrm{R,CO}}(v)$ is the radiation temperature for \element[][12]CO (cf. SMLM).
 
\begin{figure}[h!]
\centering
        \includegraphics[width=\hsize]{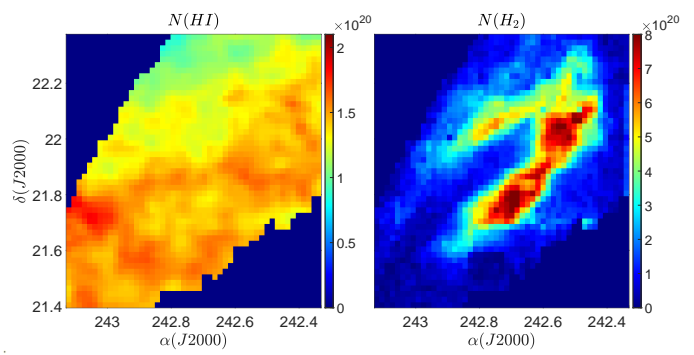}
    \caption{Total column density of  \ion{H}{i} (left panel) and H$_2$ (right panel). Outside the jagged boarders visible via N(\ion{H}{i}), the column densities were not calculated due to a lack of \element[][12]CO emission. We note that near position A, there is a deficit in atomic hydrogen where an enhancement in \element[][12]CO is present.}
        \label{fig:nh1_nh2}
\end{figure}

Fig. \ref{fig:nh1_nh2} shows the derived column densities  for  atomic and molecular hydrogen. Using equations \ref{eq:nhi} and \ref{eq:nh2}, we obtain 
$N(\mathrm{H_{tot}}) = 2N(\mathrm{H_2})+N(\ion{H}{i})$ within the cloud boundary.

\begin{figure}[h!]
\centering
        \includegraphics[width=\hsize]{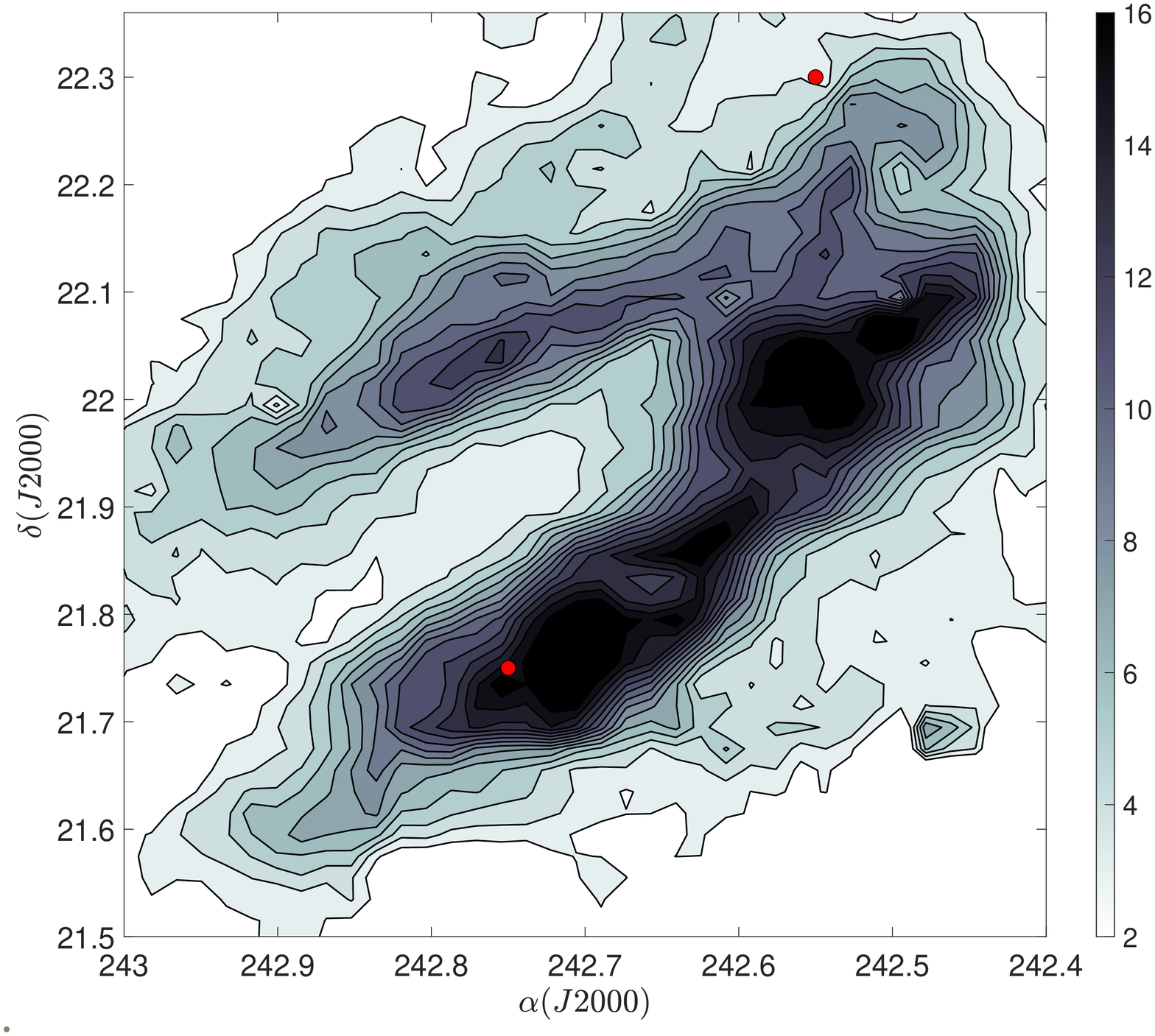}
    \caption{Total hydrogen column density. The colorbar values must be multiplied by a factor of $10^{20}$. Red dots indicate the positions discussed in Fig. \ref{fig:positions}.}
        \label{fig:total_h}
\end{figure}

Fig. \ref{fig:total_h} shows N(H$_{tot}$). The enhancement is quite steep inside the cloud, where the total column density reaches  $N(\mathrm{H_{tot}}) \approx 16 \cdot 10^{20} \ \mathrm{cm^{-2}}$. We assumed 93 pc as the distance of the cloud  (\cite{zucker2019}).  Thus,  the spatial separation between diffuse gas and the maximum hydrogen column density near position A is  $\approx 0.18 \; pc$, similar to that in \element[][12]CO, where the \element[][12]CO fades out more or less ten beams away. The cloud shows a broad atomic gas environment within the $N(\mathrm{H_{tot}}) = 2 \cdot 10^{20} \ \mathrm{cm^{-2}}$ contour (i.e., the outermost contour in Fig. 6) where the \element[][12]CO is too faint to be detected with reasonable integration times. The distribution of atomic gas is consistent with \citet{shore2003}, where it was modeled with a "cocoon" shape, but using higher fidelity from GALFA revealed some internal structure. Position B (the red dot on the top of Fig. \ref{fig:total_h}) shows \element[][12]CO weak lines with virtually no \element[][13]CO.

\subsection{Dust-to-gas mixing ratio (DGMR)}\label{subsec:dgr}

The DGMR was obtained using the IRIS 100 $\mu$m image. We linearly interpolated the image to obtain a $50 \times 50$ matrix, in which each pixel has a relative \ion{H}{i} and \element[][12]CO spectra, so we were able to directly perform a simple division to obtain the DGMR. The mean  100 $\mu$m of the MBM 40 complex is a few $\mathrm{MJy \ sr^{-1}}$ and the hydrogen column density is about $10^{20} \ \mathrm{cm^{-2}}$, so we scaled the latter by $10^{-20}$ to obtain comparable values with dust emission and then used the following ratio:
\begin{equation}
\mathrm{DGMR} = \frac{E_{100}}{N(\mathrm{H_{tot}})} \; 10^{20} \left(\mathrm{MJy \ sr^{-1} \; cm^2} \right)
,\end{equation}
where $E_{100}$ is the total emission of the IRIS 100 $\mu$m image and $N(\mathrm{H_{tot}})$ is the total hydrogen column density.

\begin{figure}[h!]
\centering
        \includegraphics[width=\hsize]{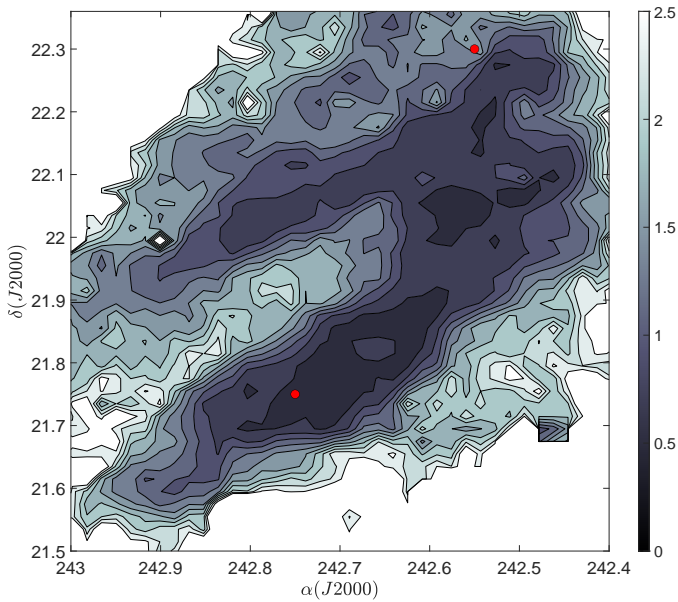}
    \caption{Map of DGMR. The colorbar values are expressed in $10^{20} \ \mathrm{MJy \ sr^{-1} \; cm^2}$. Red dots indicate the same positions as reported in Fig. \ref{fig:positions}.}
        \label{fig:dust_gas}
\end{figure}

\begin{figure}[h!]
\centering
        \includegraphics[width=\hsize]{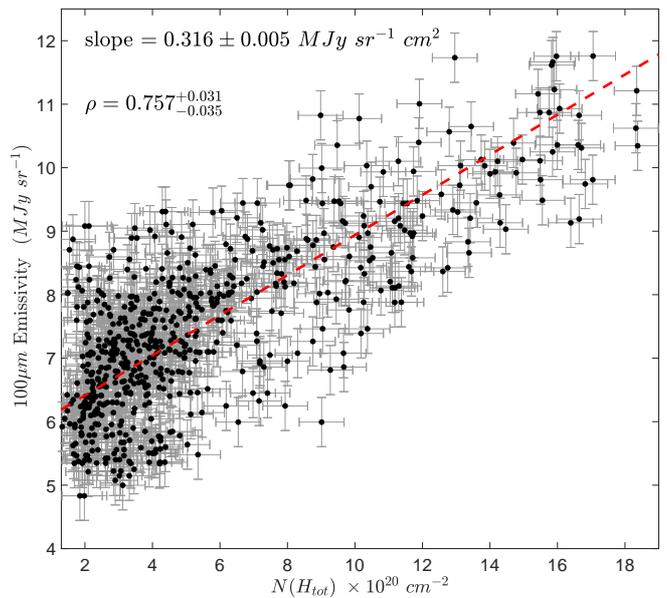}
    \caption{Scatter plot of the total column density versus 100 $\mu$m emissivity showing a linear relation. The red dashed line is the best fit with a slope of $0.316 \pm 0.005 \ MJy \ sr^{-2} \ cm^2$ and a correlation coefficient $\rho = 0.757^{+0.031}_{-0.035}$. Errors were evaluated by the standard deviation of the background where no signal is present in either $N(H_{tot})$ or 100 $\mu$m emissivity images.}
        \label{fig:correlation}
\end{figure}

The derived ratio (Fig. \ref{fig:dust_gas}) is nearly constant within the complex, indicating that the dust and gas are well mixed. We tested the robustness of our procedure by calculating the DGMR using the complete \ion{H}{i} spectra, that is  without  Gaussian decomposition and including the atomic gas over all velocities. The result is nearly the same because the majority of gas is in molecular form (mapped by \element[][12]CO). Considering the full \ion{H}{i} velocity range, the N(\ion{H}{i}) is nearly doubled, but it accounts for about 30\% of the total gas. However, without any Gaussian decomposition, the result is much more sensitive to the $X_\mathrm{CO}$ factor: if we use a lower value, that is to say 0.6 (instead of 1.3), the DGMR changes abruptly and the MBM 40 cloud is barely visible. Conversely, if we decompose the \ion{H}{i} profiles, the DGMR remains constant and also the cloud structure is clearly visible. Therefore, we can conclude that not all \ion{H}{i} gas mapped by GALFA is linked with MBM 40, but only the narrow component that we extracted.  The qualitative appearance from the map in Fig. \ref{fig:dust_gas} is quantified by the linearity of the relation of the scatterplot of the total gas column density versus 100 $\mu$m emissivity (see Fig. \ref{fig:correlation}). 

%----------------------------------------------------------------------
\section{Discussion and conclusions}\label{sec:disc}
We have shown that the dust is well mixed with gas within the translucent cloud MBM 40, both in densest and rarefied regions, indicating that HLMCs similar to MBM 40 are ideal comparisons for modelling how dust affects gas condensation.
\cite{liseau2015gas} used a different molecule, N$_2$H$^+$ instead of CO which they argued would be more condensed, to study the dust-to-gas mass ratio in $\rho$ Oph, a dense star-forming region.  In contrast, we are only concerned with the degree to which the atomic and molecular gas, and the dust, are mixed within this region.  Consequently, we do not need to make any assumptions  about the intrinsic dust properties, only that the dust temperature is approximately constant across the cloud. This is valid  for MBM 40, as confirmed by the Planck dust temperature map. 

\cite{murray2021mach} provide a detailed picture of the neutral hydrogen column densities and optical depths of diffuse gas at high Galactic latitudes.  Their mean value for N(\ion{H}{i}), around $2\times 10^{20}$cm$^{-2}$, is  similar to the highest values we have for the cocoon of MBM 40. In contrast, within the cloud boundaries, the neutral hydrogen is depleted relative to inferred H$_2$ and there we find a total hydrogen column density an order of magnitude greater (Fig. \ref{fig:total_h}), similar to the much coarser result presented in SMLM.  The maximum in N(H$_{tot}$) corresponds to the minimum in N(\ion{H}{i}).

As Murray et al. found for their sample, the gas in MBM 40  is not associated with any large  structure, such as a supernova remnant or bubble, but it appears to be a  transition to molecular gas within a more extended neutral hydrogen filamentary shear flow.  The cloud-associated atomic gas is connected in space and radial velocity to more extended regions at distances up to 10 pc in which there is no evidence for excess IRIS 100 $\mu$m emission, even in those locations where N(\ion{H}{i}) is about the same as for MBM 40.  

The comparatively small scale of the molecular structures  in MBM 40  provides a testbed for understanding the phase transition of the diffuse gas in isolated environments. The  gas and dust are well mixed regardless of the total gas density. This result should inform turbulent mixing simulations and studies of grain chemistry. Our next paper, which is currently in preparation, will present the dynamical and astrochemical tracers.

 %---------------------------------------------------------------------
\begin{acknowledgements}
\parindent 0in
We thank the Arecibo William E. Gordon Observatory staff and GALFA team for HI 21cm datacube.  The \element[][12]CO FCRAO data are obtained from SMLM 2003 study.  IRIS infrared images are obtained using the SkyView Virtual Observatory and NASA/IPAC Infrared Science Archive.   We also thank the referee for valuable suggestions that extended the discussion.
\end{acknowledgements}

% WARNING
%-------------------------------------------------------------------
% Please note that we have included the references to the file aa.dem in
% order to compile it, but we ask you to:
%
% - use BibTeX with the regular commands:
   \bibliographystyle{aa} % style aa.bst
   \bibliography{sample} % your references Yourfile.bib
%
% - join the .bib files when you upload your source files
%-------------------------------------------------------------------

\section*{Appendix: Analysis based the {\it Planck} dust maps}
 
\subsection*{Dust-to-gas mass ratio from Planck }

\begin{figure}[h!]
\centering
        \includegraphics[width=0.9\hsize]{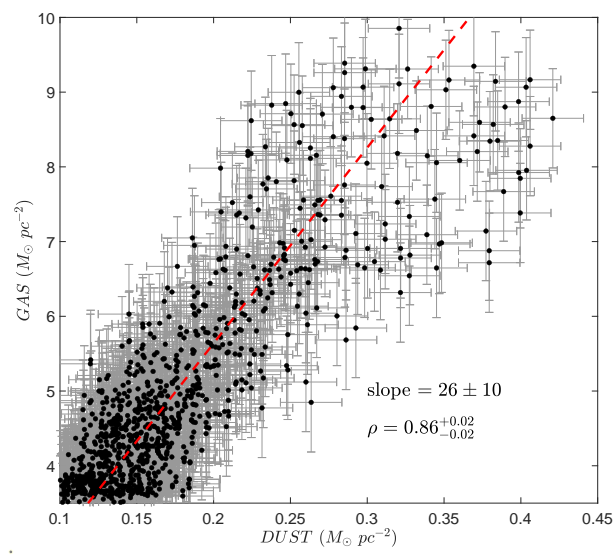}
    \caption{DGMR using Planck data.}
        \label{fig:dgr_planck}
\end{figure}

The Planck Legacy Archive (PLA) \footnote{Based on observations obtained with Planck (http://www.esa.int/Planck), an ESA science mission with instruments and contributions directly funded by ESA Member States, NASA, and Canada.} provides  high-level maps, including the dust mass column density.  We  extracted  a dust map for MBM 40 in M$_\odot$ pc$^{-2}$ and converted  our total gas column density to \ref{subsec:columnden} mass units to obtain the DGMR  (\cite{2020A&A...641A...4P}). The result is shown in Fig. \ref{fig:dgr_planck}.

%{\bf [STEVE' COMMENT]
%\newline
%I think we could leave this out, just noting that the spread is, in  part, an artifact of the reduced resolution induced by interpolation.  The more serious  problem applies to any dust data, in this case the temperature per pixel characterizes the line of sight so the SED for each and is modeled to give a unique temperature.  That's wrong  since the dust properties are as integrated as the velocities.  For isolated clouds, like the Pipe, this works far better, especially when at large distance, but for MBM 40 it's not likely more than just another way of writing en emissivity]} 

% ------- actual paragraph [MARCO] ------------------%
There is also a DGMR correlation.  The greater dispersion results from the reduced resolution caused by coordinate interpolation. But, in addition, the derivation of a single temperature for each pixel affects the correlation isolated to within the cloud.  Each pixel was integrated over the line of sight and the spectral energy distribution was modeled for each of these to give a unique temperature, but we also know from the velocity analysis that there is extended foreground and background cold dust that is not associated with MBM 40. We therefore used only the dust emission from the 100 $\mu$m IRAS image, without any assumption as to temperature modeling, and we discuss only the mixing ratio.
%----------------------------------------------------%

%----- previous paragraph [MARCO] --------------------%
%{\bf There are, however, some cautions. The Planck map has a slightly lower resolution than to data for the gas.  Furthermore, for the comparison with the gas, because it is projected in Galactic coordinates, it requires interpolation   to  equatorial coordinates that further reduces the  resolution. Finally, the Planck maps are integrated maps, so there is no spectral information; interpolating integrated maps might produce artifacts, particularly at lower resolution where using spectral data allows to perform a more robust interpolation. Third, the lower part on the left of the plot is naturally dominated by noise, where in the upper right corner the data spreads out, adducing a weak linear fit with $\chi^2 \sim 3.82 \cdot 10^4$.}
%-----------------------------------------------------%

\subsection*{Dust temperature using Planck}
\begin{figure}[h!]
\centering
        \includegraphics[width=0.9\hsize]{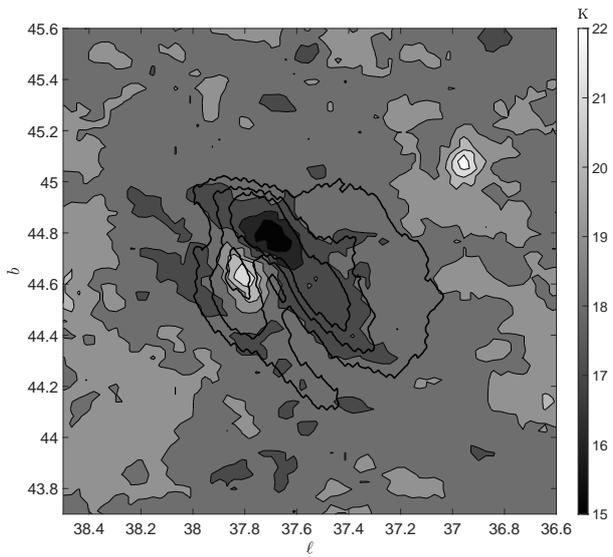}
    \caption{Dust temperature from Planck data.  Black solid contours denote MBM40 (from Planck 545 GHz channel). The two bright sources are galaxies, SDSS J161101.90+215839.6 and SDSS J160808.48+213111.7.}
        \label{fig:temperature}
\end{figure}

We selected a PLA region including MBM 40 to check whether the dust temperature is also constant through the whole cloud.  The result,  shown in Fig. \ref{fig:temperature}, is that the temperature is  nearly constant with a mean value of $\sim 18 \ K$. It is notable that the internal cloud structure is rendered uniform, confirming that the visible substructures are regions differing in density, not temperature.

\end{document}